\newif\ifpreprint
\def\preprint{\preprinttrue}

\preprint      
\ifpreprint
   \documentstyle[aps,epsf,floats,pra]{revtex} 
\twocolumn
\else
   \documentstyle[preprint,epsf,aps]{revtex}    
\fi

\newcommand {\be} {\begin{equation}} 
\newcommand {\ee} {\end{equation}} 
\newcommand {\Be}{\begin{eqnarray*}}
\newcommand {\Ee} {\end{eqnarray*}}
\newcommand {\bey} {\begin{eqnarray}} 
\newcommand {\eey} {\end{eqnarray}} 
\newcommand{\la}{\langle}
\newcommand{\ra}{\rangle}

\begin{document}
\draft

\title{Fractal dimension of space-time chaos}

\author{Antonio Politi$^{\heartsuit,\clubsuit}$ and
        Annette Witt$^{\heartsuit,\spadesuit}$
       }
\bigskip
\address{
$^\heartsuit${\it Istituto Nazionale di Ottica, L.go E. Fermi 6,
I-50125 Firenze, Italy}\\
$^\clubsuit${\it Istituto Nazionale di Fisica della Materia, Unit\`a di
Firenze}\\
$^\spadesuit${\it Nonlinear Dynamics Group, Potsdam University, \\
PF.~161535, D-14415 Potsdam, Germany}
}

\date{\today}
 
\maketitle
\begin{abstract}
A class of simplified measures is constructed to capture the key features
of generic spatio-temporally chaotic systems. A combined analytical and 
numerical investigation allows us to establish the scaling behaviour of 
the fractal dimension in open systems. Our results improve a 
previous conjecture and, what is more important, furnish a clear framework
for both numerical and analytical checks of the underlying assumptions.

\end{abstract}
\vskip 1 true cm

The structure of the invariant measure in generic spatio-temporal chaotic
systems is still far from being understood. The only basic conclusion that
is undoubtedly accepted within the physicist community is the extensivity
of the fractal dimension, i.e. that the number of active degrees of freedom
is proportional to the system size \cite{grassberger}. The number of degrees 
of freedom per unit volume is the so-called dimension density, a quantity
that can be derived from the Lyapunov spectrum, through the well known
Kaplan-Yorke formula.

However, as soon as finite subsets of (in principle) infinite systems are 
considered (i.e., an open-system point of view is adopted), it is immediately 
far from obvious how to characterize the probability distribution of generic 
observables. The common belief is that by looking at
the system with a sufficiently coarse-grained resolution, one is not able to
distinguish between a closed and an open system \cite{grassberger,politi}.
The stochastic-like action of the external world (i.e. the rest of the chain)
which activates otherwise stable degrees can be resolved only if we look at
the dynamical system with a sufficiently high resolution. This was first
suggested by Pomeau \cite{pomeau}, who postulated that the effect of distant
(in real space) degrees of freedom on the dynamics in a specific site
decreases exponentially fast with the distance. However, even though this
statement looks rather plausible, no convincing argument has yet been
presented, which justifies this hypothesis in terms of the actual behaviour
of perturbations.

The same problem has been considered in the spirit of time-series analysis
\cite{grassberger,bauer,tsimring}. In particular, Tsimring \cite{tsimring}
formulated a conjecture about the amount of information contained in finite
samples of temporal signals generated by spatio-temporal chaotic regimes. In
fact, this problem logically descends from the understanding of the
structure of the invariant measure in open systems, since the behavior of a
local observable over a finite time can be obtained by applying the
evolution operator to stationary spatial configurations in a finite region
(the so-called light cone). For this reason, we shall restrict our
discussion to this latter context in which Korzinov and Rabinovich
\cite{korzinov} have proposed a conjecture similar to those in
in Refs.~\cite{grassberger,tsimring}. Specifically, they claim that the
effective (i.e. finite scale) dimension $D_e$ in subsystems of length $L$
(for the sake of simplicity, we consider one-dimensional systems) depends on
$L$ and on the observational scale $\varepsilon$ as
\be
  D_e = dL - \frac{vd^2}{\eta} \ln \varepsilon - A ,
\label{korz}
\ee
where $d$ is the dimension density, $\eta$ is the Kolmogorov-Sinai entropy 
density \cite{grassberger}, $v$ is the propagation velocity of disturbances,
while $A$ is a non-better-specified parameter. Unfortunately, the derivation 
of the above formula is based on several assumptions that cannot be directly 
checked. Moreover, it is rather unlikely that more accurate numerical
simulations will ever provide data clean enough to draw definite conclusions.
It is therefore compelling to make some progress on the theoretical side
even at the expense of introducing strong simplifications. This is the route
already undertaken in Ref.~\cite{kantz}, where the limit case of weakly 
coupled maps has been considered. Here, rather than attempting to prove that 
the invariant measure of some dynamical system has a given structure, we have 
preferred to construct a class of measures fulfilling the basic requirements 
for a space-time chaotic regime and yet are simple enough to be handled and 
controlled. The key approximation
consists in assuming that the support of the probability distribution is a 
linear subspace, so that we can use a global approach such as singular-value 
decomposition to get at once information on the structure of the invariant 
measure on all possible length scales. Singular-value decomposition as a tool 
for analysing space-time chaos has been already profitably applied to 
experimental \cite{ciliberto} as well as to numerical \cite{greenside} data, 
allowing to identify the relevant modes. However, the validity of these 
results is limited by the unavoidable presence of nonlinearities which
definitely induce a bending of the support as well as local nonuniformities
in phase-space. Here, by referring to suitable linear subspaces, we
automatically get rid of these effects and can use a global methodology to 
extract local information. 

We shall consider a scalar process $x(i)$ defined on a lattice of length $N$,
with the label $i$ denoting the spatial position. A generic configuration
is constructed as the linear combination of some modes
\be
 x(i) = \sum_{k=1}^D p_k e_k(i) ,
\label{def}
\ee
where the $p_k$'s are i.i.d. random variables and $e_k(i)$ denotes the $i$th
component of the $k$th mode. In other words, the coefficients $p_k$ 
represent the coordinates of the process $x(i)$ in the basis of modes $e_k$.
The number of modes $D=dN$ can be read as the ``fractal'' dimension of the 
measure, while $d$ is the dimension density which is assumed to be independent
of the length, i.e. an intrinsic characteristic of the underlying 
``dynamical system''.

For the sake of simplicity, we first restrict our analysis to Fourier modes,
\be
 e_k(i)=\left\{ \begin{array}{l l}
        1/\sqrt{N} &\mbox{ if}\quad k=1\\
        \sqrt{2/N} \cos(\alpha_k i) &\mbox{ if}\quad k>1 \quad  
					  \mbox{ and odd} \\
        \sqrt{2/N} \sin(\alpha_k i) &\mbox{ if}\quad k \mbox{ even,}
          \end{array}
\right.
\label{fourier}
\ee
where $\alpha_k=2 \pi [k/2]/N$ and $[\cdot]$ denotes the integer part.

Even if very simple, such a hypothesis is not simplistic. In fact, let us 
recall that Fourier modes are the stable/unstable directions in a chain of
Bernoulli maps
$x_{t+1}(i) = {\rm Mod}[a(\mu x_t(i-1)+(1-2\mu) x_t(i)+\mu x_t(i+1)),1]$.
If one chooses the local slope $a$ and the coupling strength $\mu$  such
that the Kaplan-Yorke dimension density\cite{grassberger} is equal to $d$,
the active degrees of freedom correspond precisely to the above mentioned
$D$ Fourier modes.

The problem we want to address concerns the structure of the projection
$S^N(L,d)$ of the global invariant measure onto the lower-dimensional space
corresponding to a sub-chain of length $L \ll N$. We expect that the
resulting distribution extends along all $L$ directions. i.e. that its
dimension coincides with the space dimension $L$. Nevertheless, we
also expect $S^N(L,d)$ to be very thin along those directions corresponding
(in the closed system) to the most contracting directions (i.e., those modes
${\vec e}_k$ with $k>D$\cite{note2}).

As we are dealing with a linear subspace, the extension of $S^N(L,d)$
along the various directions can be obtained by means of the standard 
orthogonal decomposition of the correlation matrix 
\be
C_{i j}^{N}(L,d) \equiv \la x(i), x(j)\ra = 
\sum_{k=1}^D \la p_k^2 \ra e_k(i) e_k(j),
\label{equ_kovec}
\ee
where $\la \cdot \ra$ denotes an ensemble average and the last equality
can be easily proved by substituting Eq.~(\ref{def}). It is convenient
to assume that the $p_k$'s are uniformly distributed within the interval 
$[-3^{1/3},3^{1/3}]$ so that $\la p_k^2\ra=\int_{0}^{3^{1/3}}x^2 dx=1$. 
The eigenvalues $\lambda^2_l$ ($1\leq l\leq L$) of the matrix $C^N(L,d)$
represent the mean square extensions of $S^N(L,d)$ along the axes defined
by the corresponding eigenvectors.

It is possible to obtain an analytic expression for the correlation matrix.
By substituting Eq.~(\ref{fourier}) into Eq.~(\ref{equ_kovec}), and assuming 
that $D$ is odd, we find, after some simple manipulations, that
\be
C_{i j}^{N}(L,d) = \frac {1}{N}+\frac{2}{N} 
\sum_{k=1}^{(D-1)/2} \cos\Bigl((i-j) \alpha_{2k}\Bigr).
\ee 
It is immediately seen that $C_{i i}=D/N =d$, while a general expression
for the other entries can be obtained by using the relation
$\sum_{k=1}^n \cos(k x)=\cos\Bigl((n+1) x/2\Bigr) \sin(n x /2) /\sin(x/2)$ 
and other simple trigonometric formulas, 
\be
 C_{i j}^{N}(L,d)  =  \frac {\sin\Bigl( (i-j) \pi d \Bigr)}
        {N \sin\Bigl((i-j) \pi /N \Bigr)} .
\ee
In the limit $N \to \infty$ (i.e. for an infinitely extended system) but
fixed dimension density $d$ and observational length $L$, the correlation
matrix converges to
\be
 C_{i j}(L,d) =   \frac{ \sin\Bigl((i-j) d \pi\Bigr)}{\pi (i-j)}.
\label{corr_coeff}
\ee 
Similar calculations for even $D$ show that $C_{i j}$ converges to the
same limit, which thus holds in full generality.

Since the matrix entries depend only on $|i-j|$, the values along all
diagonals are constant (Toeplitz structure) and the correlation matrix
is completely determined by its first row or column. Although this structure
is rather simple, all attempts to diagonalise analytically the correlation
matrix failed. Therefore, we have been obliged to perform numerical
investigations with high precision ($10^{-32}$).

The most convenient way to look at the results is by ordering the eigenvalues
$\lambda^2_i$ from the largest to the smallest one. The qualitative behaviour
for $d=0.5$ and $L=20$ can be seen in Fig.~1, where we have reported the 
linear extensions $\lambda_i$ of the projected measure along the 
orthogonal axes represented by the eigenvectors of the correlation matrix.
We see that approximately 10 values are close to 1, while the others are
almost negligible. This confirms that, on a coarse-grained scale, an open
system looks like a closed system with the same dimension density $d$.
Nevertheless, the eigenvalues beyond the 10th are not equal to zero: this
is the result of the coupling with the pseudo-random process generated
by the external chain.

From the viewpoint of ``fractal''-dimension analysis, we can interpret these
results by saying that the effective dimension corresponding to a fixed
observational resolution $\varepsilon = lambda_i$, is $D_e = i$.
The expression for the effective dimension can thus be obtained by
inverting the expression for the eigenvalue distribution,
\be
   D_{e}(\varepsilon) = i({\lambda}) .
\ee
In the specific case reported in Fig.~1, the dimension seen for
$\varepsilon \simeq 0.1$ is still smaller than 13. Only for
$\varepsilon$-values as tiny as $10^{-7}$, the full space dimension (20) is
recovered.

This problem has a meaningful interpretation also in the context of
linear time-series analysis, since we are addressing the question of
how the average Fourier spectrum of a stochastic signal (observed in a
window of length $L$) converges to the limit shape for increasing $L$. The
difference is that here the spectrum of the signals are obtained by means of
singular value decomposition rather than finite Fourier transform.

Our main goal is to extract the relevant asymptotic behaviour of $D_e$ on
the window-length. As it can be seen in Fig.~2, where $\ln {\lambda_i}/L$
is plotted versus $i/L$ for different choices of $L$,
there is a clear evidence of an asymptotic scaling regime such as
\be
  \ln {\lambda_i} = - L F(i/L) ,
\label{scaling}
\ee
where the function $F(x)$ is identically zero for $x<d$, while it increases 
monotonously for $x>d$.
From a fundamental point of view, it is important to have a clear idea about 
the ``critical'' behaviour for $x \approx d$. In particular, it is necessary
to understand what a kind of singularity is present in that region.
The numerical analysis indicates that the right derivative $F'(x)$ computed
in $x = d$ converges to a finite value. Accordingly, for $x>d$, we can expand
$F(x)$ in a power series,
\be
  F(x) =  \sum_{j=1}^\infty \beta_j(x-d)^j
\label{scafu}
\ee
By substituting Eq.~(\ref{scafu}) in Eq.~(\ref{scaling}) and retaining only
the first two terms we find the approximate expression
\be
\ln \varepsilon \approx -\beta_1(D_e - dL) -
\frac{\beta_2}{L}(D_e - dL)^2 ,
\label{epsbeh}
\ee
where we have identified ${\lambda}$ with $\varepsilon$ and $i$ with
$D_e$. Before commenting on the physical implications of the above relation,
it is important to discuss finite-size corrections. A technical analysis
exploiting the symmetry properties of $\lambda_i^2$ suggests that the
deviation from the asymptotic value of $F(x)$ around $i = dL$ are equal to
$-(\ln 2)/2L$\cite{preprint}.

By including such corrections in Eq.~(\ref{epsbeh}) and thereby inverting, 
we find that 
\be
  D_e = dL - \frac{\ln \varepsilon}{\beta_1} -
    \frac{\beta_2}{\beta_1^3L}(\ln \varepsilon)^2  - \frac{\ln 2}{2} .
\label{deff}
\ee
The structure of this expression reduces to that of the conjecture in
Ref.~\cite{korzinov} (i.e. Eq.~(\ref{korz}) provided that the
squared logarithmic term is negligible, i.e. provided that
$L \gg \ln \varepsilon$. One can easily check that this inequality is
not always satisfied in direct computations of the dimension, so that it
would be worth re-analyzing numerical data in order to obtain reliable
estimates of the various coefficients.

A further difference with Eq.~(\ref{korz}) resides in the coefficient in
front of the logarithmic term: we find $1/\beta_1$ that is to be confronted
with $vd^2/\eta$. We are able to determine $\beta_1$ only numerically (even
though rather accurately) and it turns out to be approximately equal to 0.6
independently of $d$ in the interval $[0.2,0.8]$. The natural model to
compare our expectations with Eq.~(\ref{korz}) is again the chain of
Bernoulli maps in which case, one can determine analytically the Lyapunov
spectrum\cite{isola} and easily compute $d$, $\eta$, and $v$. Tests made for
different choices of the slope $a$ and of the diffusive coupling $\mu$ show
that $vd^2/\eta$ is always significantly larger than $1/\beta_1$
(by almost a factor 2). This inequality hints at a possible consistency of
the two approaches. In fact, Eq.~(\ref{deff}) can at least be viewed as a 
lower bound for the effective dimension, since it has been obtained by
neglecting nonlinearities (which enter as discontinuities in Bernoulli maps)
which can only contribute to increase the dimension on finite scales.
Anyhow, it would be very important to establish whether the discrepancy
is to be attributed to a failure of the conjecture in Ref.~\cite{korzinov},
or to ingredient neglected in our treatment. In favour of the former
hypothesis, we claim that, as long as nonlinearities imply only large scale
phenomena such as bendings and nonuniformities, they should not affect the
asymptotic behaviour described by the scaling function $F(x)$. Indeed, from 
the point of view of an information-theoretic approach (which is implicitely 
that one adopted by deriving an expression for the effective dimension), 
there is no difference between a straight and a bended manifold, provided 
that the bending is not too strong \cite{note1}: the same number of boxes
is needed to cover the set.

Aside from the role of nonlinearities, the validity of our results could be
challenged on the basis of the special assumptions made to construct
the invariant measure. Therefore, we have decided to progressively remove 
some of the limitations. We have started by assuming that the average 
amplitude of the first (nonzero) $D$ Fourier modes is not constant but goes 
continuously to 0. In this case, one can again find an explicit expression 
for the correlation matrix; numerical studies performed for different choices 
of the Fourier amplitudes indicate a convergence towards the same function
$F(x)$ but stronger finite-size corrections (i.e. larger $A$).

A further criticism could be that the Fourier basis is a rather special
choice, contrasting with the observation that Lyapunov vectors are
localized\cite{kane}. We have therefore considered also bases made of
exponentially localized vectors. This has been done by implicitely referring
to Bernoulli maps with quenched disorder (i.e. by randomly fixing the slope
of the local maps). In this case, the correlation matrix can only be 
constructed numerically and for lattices of finite length $N$; moreover,
there is the additional difficulty of the dependence on the realization of
the disorder. Surprisingly enough, we find again a reasonable agreement with 
the results obtained for the Fourier basis with the same system-size (provided
that the geometric average of the eigenvalue spectra is taken). Accordingly, 
we are led to conjecture that the behaviour displayed by the Fourier basis is 
quite universal and seemingly independent even of the localization properties 
of the basis.

Summarizing, we have found a rather general scaling law expressing the
dependence of the effective dimension on the size of an open system
and on the observational resolution. As the result has been derived
under the assumption of a linear structure for the invariant measure, it is
now crucial to test its validity in generic models, where nonlinearities play
certainly an important role.

AW thanks the Max-Planck-Society for financial support. A. Torcini
is acknowledged for useful discussions, while M. B\"unner, A. Pikovsky,
and M. Zaks for a careful reading of the manuscript.

\begin{figure}
\caption{Extensions along the orthogonal directions for an object with
dimension density $d=0.5$ and an embedding dimension $L=20$.}
\label{fig1} 
\end{figure}

\begin{figure}
\caption{Scaling of the eigenvalue spectrum of $C(L,d)$ for different 
spatial lengths $L$ but fixed dimension density $d=0.5$. The
increasingly large plateaus seen for large values of $L$ are a numerical
artifact signaling that the limit of computer accuracy has been
reached.}
\label{fig2} 
\end{figure}

\smallskip


\begin{thebibliography}{99}
\bibitem{note2} Let us indeed recall that in Bernoulli maps, Fourier modes
with increasing wavenumber are increasingly stable if $\mu <0.5$
(see Ref.~\cite{isola}).
\bibitem{note1} Different would be the problem of the possibly Cantor-like
structure along some directions. However, if this phenomenon is restricted 
to just one direction, as implicit in the Kaplan-Yorke formula, it should 
not imply relevant corrections to our analysis.
\end{thebibliography}
\end{document}